\documentclass[iop,floatfix]{emulateapj}

\usepackage{graphicx}
\usepackage{amsmath}
\usepackage[colorlinks=true,linkcolor=blue,citecolor=blue,urlcolor=blue]{hyperref}
\usepackage{mypack}
\usepackage{comment}

\newcommand{\srcnm}{SGR J1745\ensuremath{-}2900}
\newcommand{\nepochs}{33}

\shorttitle{GBT and \textit{Swift} Observations of \srcnm}
\shortauthors{Lynch et al.}

\begin{document}

\title{Green Bank Telescope and \textit{Swift} X-ray Telescope
  Observations of the Galactic Center Radio Magnetar \srcnm}

\author{Ryan S.\ Lynch\altaffilmark{1}, Robert F.\ Archibald, Victoria
  M.\ Kaspi, and Paul Scholz} \affil{Department of Physics, McGill
  University, 3600 University Street, Montreal, Quebec, H3A 2T8,
  Canada} \altaffiltext{1}{Department of Physics and Astronomy, West
  Virginia University, PO Box 6315, Morgantown, WV 26506, USA}
\email{rlynch@physics.mcgill.ca}

\begin{abstract}

We present results from eight months of Green Bank Telescope
$8.7$-$\GHz$ observations and nearly 18 months \textit{Swift} X-ray
telescope observations of the radio magnetar \srcnm, which is located
$2\farcs4$ from Sgr A\textsuperscript{*}.  We tracked the magnetar's
radio flux density, polarization properties, pulse profile evolution,
rotation, and single-pulse behavior.  We identified two main periods
of activity in \srcnm.  The first is characterized by approximately
5.5 months of relatively stable evolution in radio flux density,
rotation, and profile shape, while in the second these properties
varied substantially.  Specifically, a third profile component emerged
and the radio flux increased on average, but also became more
variable.  Bright single pulses are visible and are well described by
a log-normal energy distribution at low to moderate energies, but with
an excess at high energies.  The $2$--$10\; \keV$ flux has decayed
steadily since the initial X-ray outburst, in contrast with the radio
flux.  Our timing analysis includes Green Bank Telescope,
\textit{Swift}, and \textit{NuSTAR} data.  When we include the X-ray
data in our analyses, we find that \srcnm\ exhibited a level of timing
noise unprecedented in a radio magnetar, though an analysis of only
the radio data indicates timing noise at a level similar to that
observed in other radio magnetars.  We conclude that, while \srcnm\ is
similar to other radio magnetars in many regards, it differs by having
experienced a period of relative stability in the radio band that now
appears to have ended, while the X-ray properties have evolved
independently.

\end{abstract}


\section{Introduction}
\label{sec:intro}

Magnetars are neutron stars that display intense short X-ray bursts,
X-ray pulsations, and X-ray outbursts that are typically followed by a
decaying X-ray flux.  The hallmark of magnetar activity is an X-ray
luminosity that may exceed the power available from the neutron star's
rotation.  Instead, magnetars are believed to be powered by the decay
of their enormous internal magnetic fields
\citep{dt95,dt96,tlk02,bel09}, and the most active
magnetars\footnote{For an up-to-date list, see the McGill magnetar
  catalog at
  \\ \url{http://www.physics.mcgill.ca/~pulsar/magnetar/main.html}
  \citep{ok14}.}  generally have inferred surface dipolar magnetic
fields of $10^{14}$--$10^{15}\; \gauss$, much higher than the $\sim
10^{12}\; \gauss$ typical of rotation-powered radio pulsars.  Under
this interpretation, as the internal magnetic field decays, it
stresses the stellar crust, inducing occasional sudden crustal and/or
magnetospheric reconfigurations that give rise to the variety of
variable X-ray emission.

Pulsed radio emission has been detected from four magnetars thus far
\citep{crh+06,crhr07,lbb+10,sj13}.  Their radio properties show both
similarities and marked differences when compared to those of
rotation-powered pulsars.  Like all magnetars \citep[e.g.][]{dk14},
the four radio magnetars have a high degree of timing noise and
experience significant changes in torque \citep[e.g.][]{ccr+07}.  Both
rotation-powered pulsars and radio magnetars exhibit a high degree of
linear polarization \citep{crj+07,crhr07,ksj+07,lbb+12,sj13,efk+13}
and have radio spectra that can be well modeled with a single power
law, $S_\nu \propto \nu^\alpha$.  However, magnetars have shallow or
even flat spectra, with $\alpha$ that may vary significantly with time
\citep[e.g.][]{crp+07}, whereas rotation-powered pulsars typically
have stable spectra with $\langle \alpha \rangle \sim -1.6$
\citep{lylg95}.  The flux density and pulse profile morphology of some
radio magnetars are also highly variable \citep[e.g.][]{ccr+07}.  This
may be due, at least in part, to the emission of erratic, extremely
narrow single pulses and very long profile stabilization timescales
\citep{ksj+07}.  Single pulse studies of magnetars have not revealed
evidence for drifting sub-pulses of the kind sometimes seen in
rotation-powered pulsars \citep{ssw+09}.

The newest of the four known radio magnetars is
\srcnm\footnote{\citet{ok14} also refer to this source as MG
  J1745$-$2900.}.  The X-ray source was discovered in outburst by
\citet{kbk+13} using \textit{Swift}.  Subsequent observations with the
\textit{NuSTAR} X-ray telescope detected pulsations with a spin period
$P = 3.76\; \s$ and $\dot{P} = 6.5 \times 10^{-12}\; \s\, \s^{-1}$,
implying a magnetic field of strength $B = 1.6 \times 10^{14}\;
\gauss$, thus confirming the pulsar's nature as a magnetar
\citep{mgz+13}.  Radio pulsations at the same period were subsequently
detected at several observatories \citep{sj13,efk+13}.  \srcnm\ lies
only $2\farcs4$ in projection from Sgr A\textsuperscript{*}
\citep{rep+13}, and its dispersion measure ($\dm = 1778 \pm 3\; \dmu$)
implies that the source lies $< 10\; \pc$ from Sgr
A\textsuperscript{*} itself \citep{efk+13}.  Early radio observations
have measured a flat spectrum and high degree of polarized emission
for \srcnm, much like in other radio magnetars \citep{ekc+13,sj13}.

Here, we report on the results from an observing campaign of
\srcnm\ using the Robert C.\ Byrd Green Bank Telescope (GBT),
supplemented with data from the \textit{Swift} X-ray Telescope (XRT)
and \textit{NuSTAR} telescope.  We tracked the evolution of the spin,
radio flux density, polarization, and profile morphology of the pulsar
at $8.7\; \GHz$, and have also analyzed properties of its single
pulses.  XRT was used to measure the $2$--$10\; \keV$ flux.  We first
provide relevant background on the magnetar's behavior in
\S\ref{sec:overview}.  In \S\ref{sec:data} we describe our
observational set-up and data reduction.  Our analysis and results are
presented in \S\ref{sec:analysis}, and discussed in more detail in
\S\ref{sec:discussion}.

\begin{figure*}[t]
  \centering
  \includegraphics[width=0.9\textwidth]{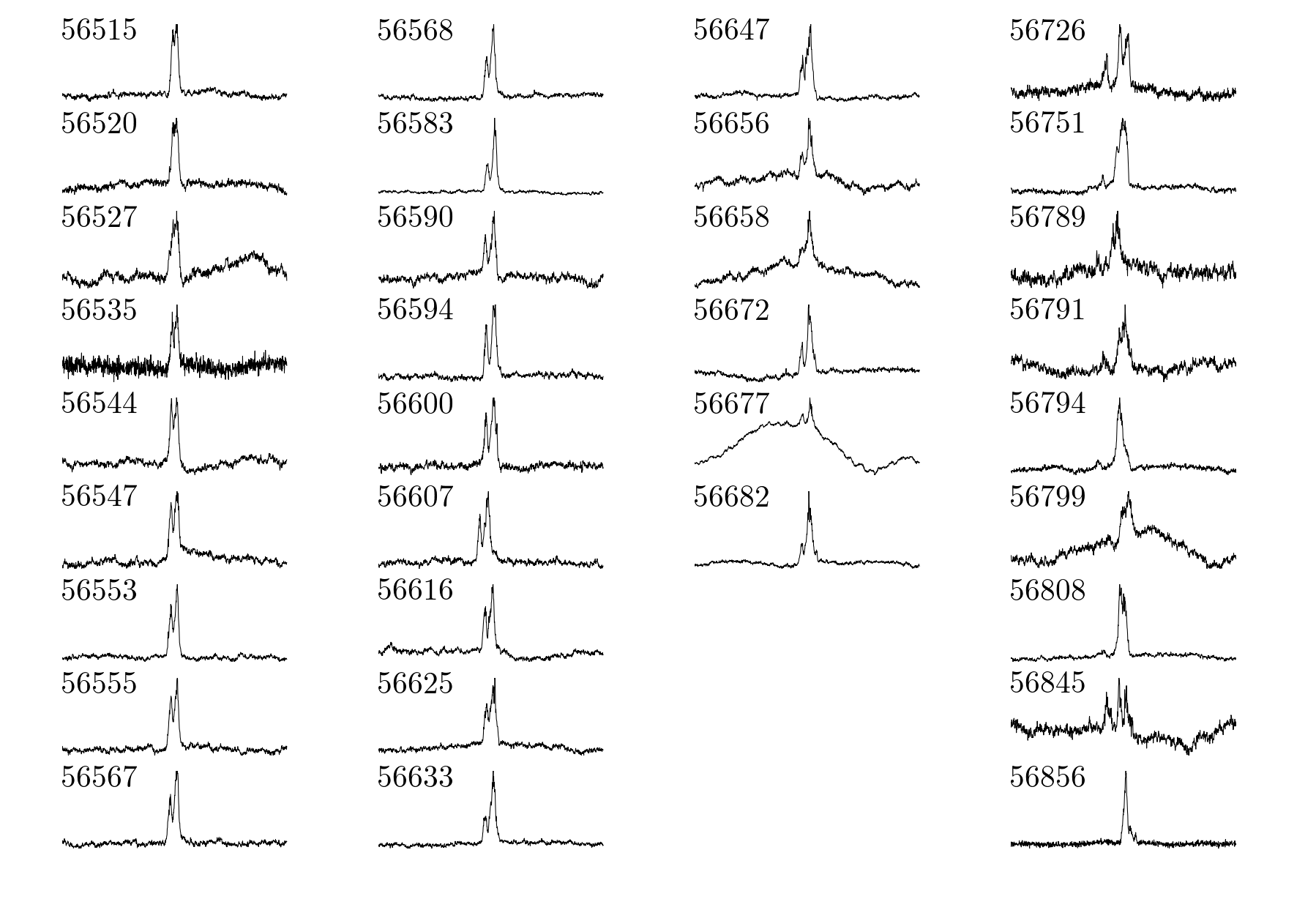}
  \caption{Integrated $8.7$-$\GHz$ pules profiles of \srcnm\ at each
    of our observing epochs, which are indicated by MJD.  For clarity,
    profiles are centered and scaled to have the same peak amplitude.
    As described in the text, baseline fluctuations that we attribute
    to changes in atmospheric opacity are evident at certain epochs,
    and are not intrinsic to \srcnm.  Up to MJD 56682 (first three
    columns), the profile shape was fairly stable, exhibiting a clear
    double peaked structure.  Changes in the relative amplitude of the
    two peaks are most likely due to the pulse-to-pulse variability,
    given the relatively small number of rotations observed at each
    epoch, which prevents the profile from fully stabilizing.
    Starting with MJD 56726 (fourth column), the profile became more
    variable, often showing a third component.  \label{fig:profiles}}
\end{figure*}

\section{Overview of Radio and X-ray Behavior}
\label{sec:overview}

Before we discuss our results in detail, we provide some context by
highlighting some of the radio and X-ray properties of
\srcnm\ reported elsewhere.  Radio pulse profiles obtained shortly
after the magnetar was first detected had a single Gaussian component
across a wide range of frequencies, including $\sim 8$--$9\; \GHz$
\citep{sj13,sle+14}, though there appears to be visual hints of a
second profile component emerging in some of the profiles presented by
\citet{efk+13}.  The early reported flux densities near $9\; \GHz$
varied but were of order $\sim 1\; \mJy$.

After the initial outburst, the X-ray flux of \srcnm\ began to decay.
The \textit{Swift} Burst Alert Telescope (BAT) detected short bursts
on MJDs 56407 \citep[25 April 2013;][]{kkb+13}, 56450 \citep[7 June
  2013;][]{kbckd+13}, and 56509 \citep[5 August 2013;][]{kbckb+13}.
Despite this, the overall flux decay continued.  \citet{kab+14}
suggested that the burst on MJD 56450 may have been accompanied by a
change in spin-down ($\dot{f}$), though there was no discontinuity in
spin frequency.  This behavior was not consistent with a glitch or
anti-glitch, though timing noise could not be ruled out.

\citet{yrh+14} observed \srcnm\ on 25 February 2014 (MJD 56713) at
$44\; \GHz$ with the Very Large Array and measured a flux density that
was $\sim 20$ times higher than an upper limit obtained in August
2011.  Simultaneous \textit{Chandra} observations showed no evidence
of a corresponding increase in X-ray flux \citep{rhb+14}.  We observed
\srcnm\ 31 days prior to and 13 days after this $44$-$\GHz$ flux
density increase (on MJDs 56682 and 56726, respectively).  As
discussed below, our data provide evidence that \srcnm\ underwent a
change in flux density, pulse profile morphology, and possibly
rotational parameters sometime between the above two observations,
changing from a fairly stable state to a more erratic radio state.  We
will refer to the stable state (covering MJDs 56515--56682) and
erratic state (covering MJDs 56726--56845) throughout.  We will
discuss the extent with which we can associate this state change with
the $44$-$\GHz$ radio brightening in \S\ref{sec:discussion}.

\section{Observations and Data Reduction}
\label{sec:data}

\subsection{Green Bank Telescope}
\label{sec:gbtobs}

We observed \srcnm\ on \nepochs\ epochs.  We used an approximately
weekly cadence from 11 August 2014 through 25 January 2014\footnote{No
  observations were possible from 4--18 October 2013 because GBT
  operations were suspended due the U.S.\ Federal Government
  shutdown.}, during which integration times averaged about 30
minutes.  Because \srcnm\ was fairly stable over this period, we
planned on switching to a monthly cadence with two hour integrations
starting in February 2014.  However, after two monthly monitoring
sessions, it became apparent that the magnetar increased in
variability.  Therefore, we re-allocated our time to allow for more
frequent but shorter, 30 minute, observations, though scheduling
constraints prevented us from restarting weekly sessions.  

We used the X-band receiver system of the GBT at a center frequency
$\nu_\mathrm{c} = 8.7\; \GHz$ and with an instantaneous bandwidth of
$\Delta \nu = 800\; \MHz$, recording dual circular polarizations.  The
data were recorded with the the Green Bank Ultimate Pulsar Processing
Instrument, using $512$ frequency channels and a sampling time of
$\delta t = 163.84\; \us$.  At the beginning of each observing session
we took on and off-source scans of a standard flux calibrator while
firing the GBT pulsed noise diode.  We initially used 3C353 as a flux
standard, but after noticing anomalies in the flux densities we
calculated, we switched to using 3C286 on and after 12 May 2014 (see
\S\ref{sec:flux} for a detailed discussion of our reported flux
density measurements).  We observed the noise diode again at the
position of \srcnm\ before observing the magnetar, and used these data
for calibration purposes.

The data were folded modulo the rotational period of \srcnm, using
$1024$ bins in pulse phase and sub-integrations with a duration of one
rotation, thus preserving information on individual single pulses.
Radio frequency interference (RFI) was usually minimal, but when
present we excised it manually by explicitly removing contaminated
frequency channels and sub-integrations.  We also removed $\sim 2.5\%$
from the top and bottom of the frequency band due to roll-off in the
receiver sensitivity.  This was usually sufficient for obtaining
integrated pulse profiles with few artifacts, but on certain epochs
significant fluctuations in the off-pulse region were evident (as an
example, see the profile from MJD 56677 in Figure \ref{fig:profiles}).
The effect is usually (though not exclusively) associated with
observations that occurred at elevation angles below $\sim 10^\circ$,
and we believe that they are primarily caused by changes in
atmospheric opacity.  To ensure that the fluctuations are not
intrinsic to \srcnm, we folded the data at a period randomly drawn
from a uniform distribution on an interval $P \pm 0.5\; \s$, where $P
= 3.76\; \s$ is the spin period of the magnetar.  This effectively
allowed us to sample the period space around the magnetar while
avoiding bias in the chosen period.  As expected, similar fluctuations
were readily apparent when folding at the randomly selected period,
demonstrating that they are extrinsic to \srcnm.

After folding and RFI excision, the total, linearly, and circularly
polarized flux densities were calibrated using standard routines from
the \texttt{PSRCHIVE} software package \citep{hvm04}.  On and
off-source observations of the standard flux calibrator were used to
measure the absolute flux density of the pulsed noise diode, and this
in turn was used to calibrate the total and polarized flux density of
\srcnm.  We corrected for Faraday rotation by searching over a range
of rotation measures (RMs, again using \texttt{PSRCHIVE} standard
tools), from $-2 \times 10^5 \leq \mathrm{RM} \leq 0\; \mathrm{rad\:
  \m^{-2}}$, de-rotating our data at the RM that maximized the
linearly polarized flux density.  All subsequent analyses were
performed on these calibrated, RM-corrected data unless otherwise
noted.

\subsection{\textit{Swift} XRT}
\label{sec:swiftobs}

In order to characterize the X-ray flux evolution of \srcnm, we
analyzed 416 \textit{Swift} XRT observations of the source obtained
between MJDs 56407 and 56956 as part of the Galactic center monitoring
program \citep{dmk+13}.  Observations were typically $1$-$\ks$ long
and occurred nearly daily, except between MJDs 56599 and 56690, when
the source was in Sun-constraint.  A total of $\sim 438\; \ks$ of data
were analyzed.

The XRT \citep{bhn+05} is a Wolter-I telescope with an
\textit{XMM-Newton} EPIC-MOS CCD22 detector, sensitive in the
$0.5$--$10\; \keV$ range.  For all the observations presented here,
the XRT was operated in Photon Counting (PC) mode, which has a time
resolution of $2.5\; \s$.  We obtained Level-1 data products from the
HEASARC \textit{Swift} archive, reduced them using the
\emph{xrtpipeline} standard reduction script, and reduced them to the
Solar system barycenter them using the \textit{Chandra} position
\citep{rep+13} of \srcnm\, using HEASOFT v6.16.  Individual exposure
maps, spectra, and ancillary response files were created for each
orbit and then summed.

\begin{figure*}[t]
  \centering
  \includegraphics[width=\textwidth]{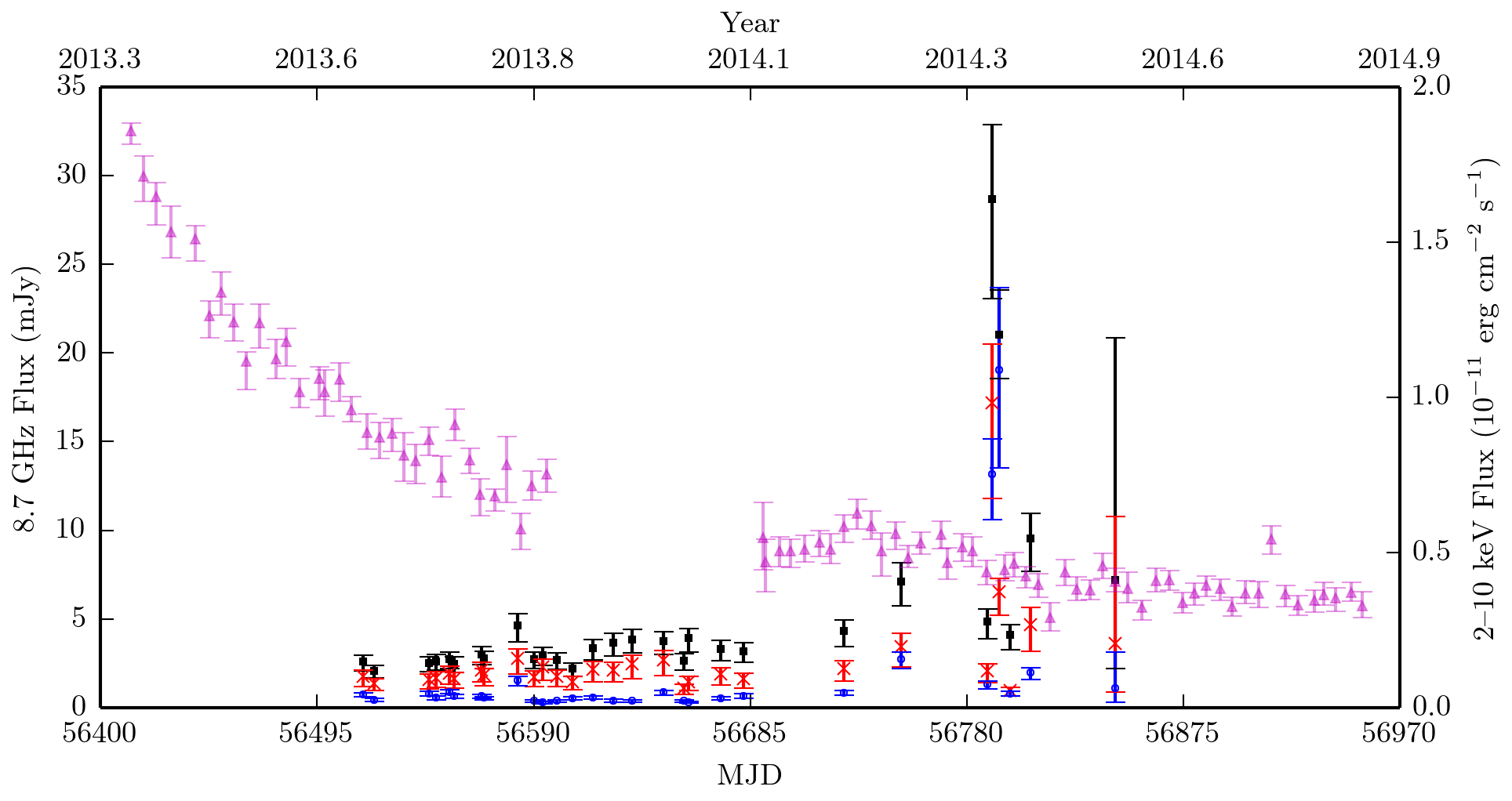}
  \caption{The period averaged $8.7$-$\GHz$ flux density and
    $2$--$10\; \keV$ flux of \srcnm.  The squares (black), crosses
    (red), and circles (blue) indicate total, linearly polarized, and
    circularly polarized radio flux density, respectively.  X-ray
    fluxes are indicated by triangles (magenta).  Within the
    uncertainties in our measurements, the radio flux was relatively
    stable up to MJD 56682, with a mean of $\sim 3\; \mJy$.  After MJD
    56726, the flux was more erratic.  The X-ray flux has decayed
    steadily since the initial outburst and is well modeled by a
    double exponential function (see text for parameters).  The
    moderate X-ray flux increase around MJD 56731 is associated with
    another source.  The flare on MJD 56910 is most likely associated
    with Sgr A\textsuperscript{*}.  \label{fig:fluxes}}
\end{figure*}

\section{Analysis}
\label{sec:analysis}

\subsection{Radio Flux and Polarization Properties}
\label{sec:flux}

We began the process of measuring the mean flux density, $S_\nu$, of
\srcnm\ by systematically identifying the on-pulse region at each
epoch.  This was done by identifying the profile bins to both the left
and right of the profile peak at which the flux density reached the
root-mean-square (RMS) level of the integrated profile as a whole.
This gave us a rough idea of the on-pulse and off-pulse bins.  We
iterated this procedure using the RMS of the updated off-pulse region
until the results converged.  The baseline fluctuations described in
\S\ref{sec:data} may bias this procedure in two ways---by raising the
off-pulse RMS and by making the on-pulse region appear broader than it
actually is.  We attempted to mitigate this by fitting a third order
polynomial to the integrated profile (but using only the off-pulse
region) and subtracting it to flatten the baseline.  We used this
approach to determine the on-pulse region, but used the the original,
unflattened profile to calculate $S_\nu$.

We found that, when using 3C353 as flux calibrator, our calculated
$8.7$-$\GHz$ flux densities were consistently a factor of $\sim
10$--$20$ higher than those reported by \citet{sj13} and
\citet{efk+13} at similar frequencies.  To confirm this discrepancy we
observed both 3C353 and 3C286 on the same date, and calibrated our
data from that date independently using both sources.  The data
calibrated using 3C286 were in rough agreement with the two previous
reports of $S_\mathrm{8.7\; GHz}$ (which we note were obtained
independently from each other with different telescopes), while the
data calibrated using 3C353 were a factor of several higher.  From
this, we concluded that 3C353 is not a reliable flux standard at these
frequencies.  We subsequently used only 3C286 for flux calibration,
observing it at a total of five epochs, though a hardware error made
one of these unusable for flux calibration.  To measure the flux
density at all of our observing epochs, we independently calibrated
all of our data using the four reliable 3C286 observations.  This
resulted in four separate flux measurements at each of our
\nepochs\ observing epochs, although three epochs had to be discarded
from the flux density analysis because of malfunctions of the noise
diode.  Even though we used the same flux standard, there was still
scatter among these four flux density values.  The flux density we
report here is the mean of these and the uncertainties represent the
minimum and maximum of the four independent flux density values.  Our
experience highlights the inherent difficultly of obtaining reliable
absolute flux density measurements and can hopefully serve as a
cautionary tale to other observers.

\begin{figure}[b]
  \centering
  \includegraphics[width=\columnwidth]{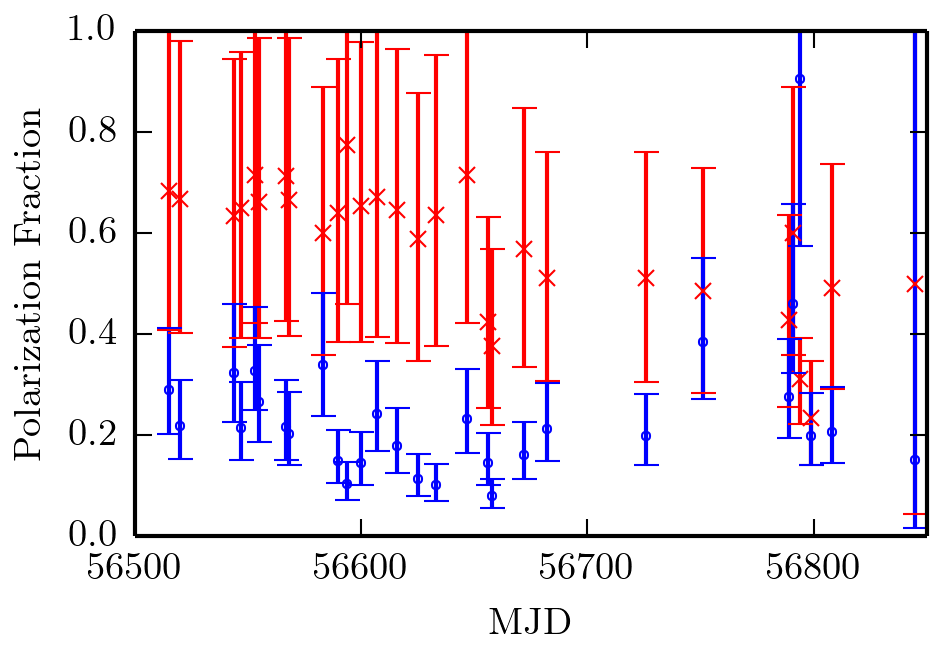}
  \caption{The fraction of linearly and circular polarization.  The
    symbols are the same as in the top panel.  During the stable state
    the linear polarization fraction was relatively constant at $\sim
    0.6$, though the uncertainties are large.  The circular
    polarization fraction was typically $\sim 0.2$. During the erratic
    state, the circular polarization fraction increased to a mean of
    $\sim 0.4$ while the linear polarization fraction decreased to a
    mean of $\sim 0.4$.  \label{fig:pols}}
\end{figure}

The radio flux density is shown in \ref{fig:fluxes} (along with the
X-ray flux), while the fractional polarization is shown in
\ref{fig:pols}.  During the previously identified stable state (see
\S\ref{sec:overview}), the flux density of \srcnm\ was nearly
constant, with a mean $S_{\rm 8.7\; GHz} \approx 3.0\; \mJy$ and a
standard deviation of $0.62\; \mJy$ between measurements.  For
comparison, the mean uncertainty in our measurements of $S_\mathrm{
  8.7\; GHz}$ during this state was $^{+0.6}_{-0.4}\; \mJy$.  Like
other authors \citep{sj13,efk+13}, we also measure a high degree of
linear polarization.  The mean linear polarization fraction in the
stable state was $0.6$ with a standard deviation of $0.09$.  The
circular polariziation was much smaller at $0.2$ and with a standard
deviation of $0.08$.

During the erratic state, the flux of \srcnm\ was both higher and more
variable.  The mean flux was $S_\mathrm{8.7\; GHz} \approx 11\; \mJy$
but with a standard deviation of $8.5\; \mJy$ and mean uncertainties
of $^{+3.1}_{-2.4}\; \mJy$.  Not only was the standard deviation in
$S_\mathrm{8.7\; GHz}$ a substantial fraction of the mean ($0.78$), it
was much larger than the uncertainties.  The circular polarization
fraction increased slightly, with a mean of $0.4$ and standard
deviation of $0.2$, while the linear polarization fraction decreased
to a mean of $0.4$ with standard deviation of $0.1$.

\subsection{X-ray Flux}

To investigate the X-ray flux and spectral behavior, we extracted a
circular region centered on \srcnm\ and with a diameter of a
$20\arcsec$, chosen to match the half-power diameter of the XRT at
$4\; \keV$.  An annulus of inner radius $20\arcsec$, and outer radius
$60\arcsec$ centered on the source was used to extract background
spectra.  This is the same background region used by \citet{kbk+13}.
As the angular distance between \srcnm\ and Sgr A\textsuperscript{*}
is $2\farcs4 \pm 0\farcs3$, we note that our source region also
contains Sgr A\textsuperscript{*}.  We summed the spectra in five-day
intervals, and grouped them to have a minimum of three counts per bin.
Photoelectric absorption was modeled using \texttt{XSPEC}
\texttt{tbabs}, with abundances from \citet{wam00}, and photoelectric
cross-sections from \citet{vfky96}.  We then fit the spectra to a
photoelectrically absorbed black body using the ``lstat'' statistic.
The spectra were fit jointly, with a single neutral hydrogen column
density ($N_{\rm H}$) and temperature ($kT$), allowing only the black
body normalization to vary for each spectrum.  A variable $kT$ was not
statistically warranted.  This gave best-fit values of $N_{\rm H} =
(12.1 \pm 0.3) \times 10^{22}\; \cm^{-2}$ and $kT = 1.00 \pm 0.01 \;
\keV$ ($\chi^2 = 8585$ for $8917$ degrees of freedom, or
$\mathrm{lstat} = 8620.68$). These values of $N_{\rm H}$ and $kT$ are
consistent with those reported by \citet{kbk+13}.

The evolution of $2$--$10\; \keV$ flux is shown alongside $S_{\rm
  8.7\; GHz}$ in Figure \ref{fig:fluxes}.  The flux decay is
reasonably well fit ($\chi^2 = 92.6$ for $76$ degrees of freedom) by
the sum of two exponential decay functions:
\begin{multline}
F= \Big[ (1.00 \pm 0.06) e^{-(t-t_0)/(55 \pm 7\; \mathrm{d})} +
  \\ (0.98 \pm 0.07) e^{-(t-t_0)/(500 \pm 41\; \mathrm{d})} \Big]
\\ \times 10^{-11}\; \erg\, \cm^{-2}\, \s^{-1},
\end{multline}
where $t_0 = 56406$ is the the peak of the outburst.  There is is a
small flux increase around MJD 56731 due to nearby source leaking into
the extraction region, and is not related to the magnetar.  There is
also a flare evident on MJD 56910 that is likely due to Sgr
A\textsuperscript{*} \citep{drm+14}.  We see no significant change in
the $2$--$10\; \keV$ flux coincident with onset of erratic radio
behavior.

\subsection{Radio Profile Shape Evolution}
\label{sec:shape}

\begin{figure}[t]
  \centering \includegraphics[width=\columnwidth]{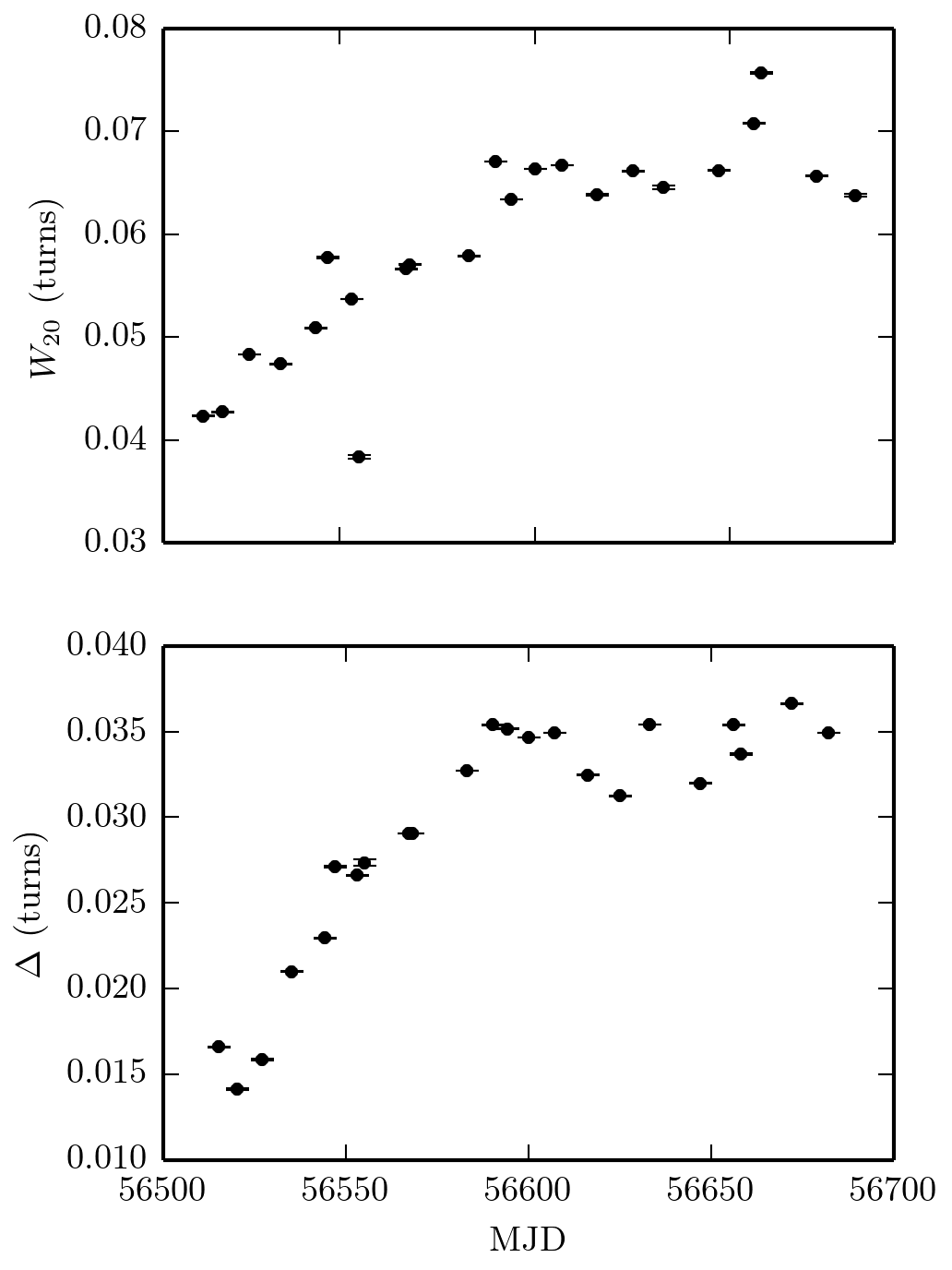}
  \caption{Full width at $20\%$ of peak flux ($W_{20}$) and
    peak-to-peak separation ($\Delta$) as a function of time for
    \srcnm.  Only the stable state is shown.  Error bars indicate the
    $1$-$\sigma$ confidence intervals.  There is a linear increase in
    both parameters between MJDs 56544 and 56594.
     \label{fig:shapes}}
\end{figure}

\begin{figure*}[t]
  \centering
  \includegraphics[width=\textwidth]{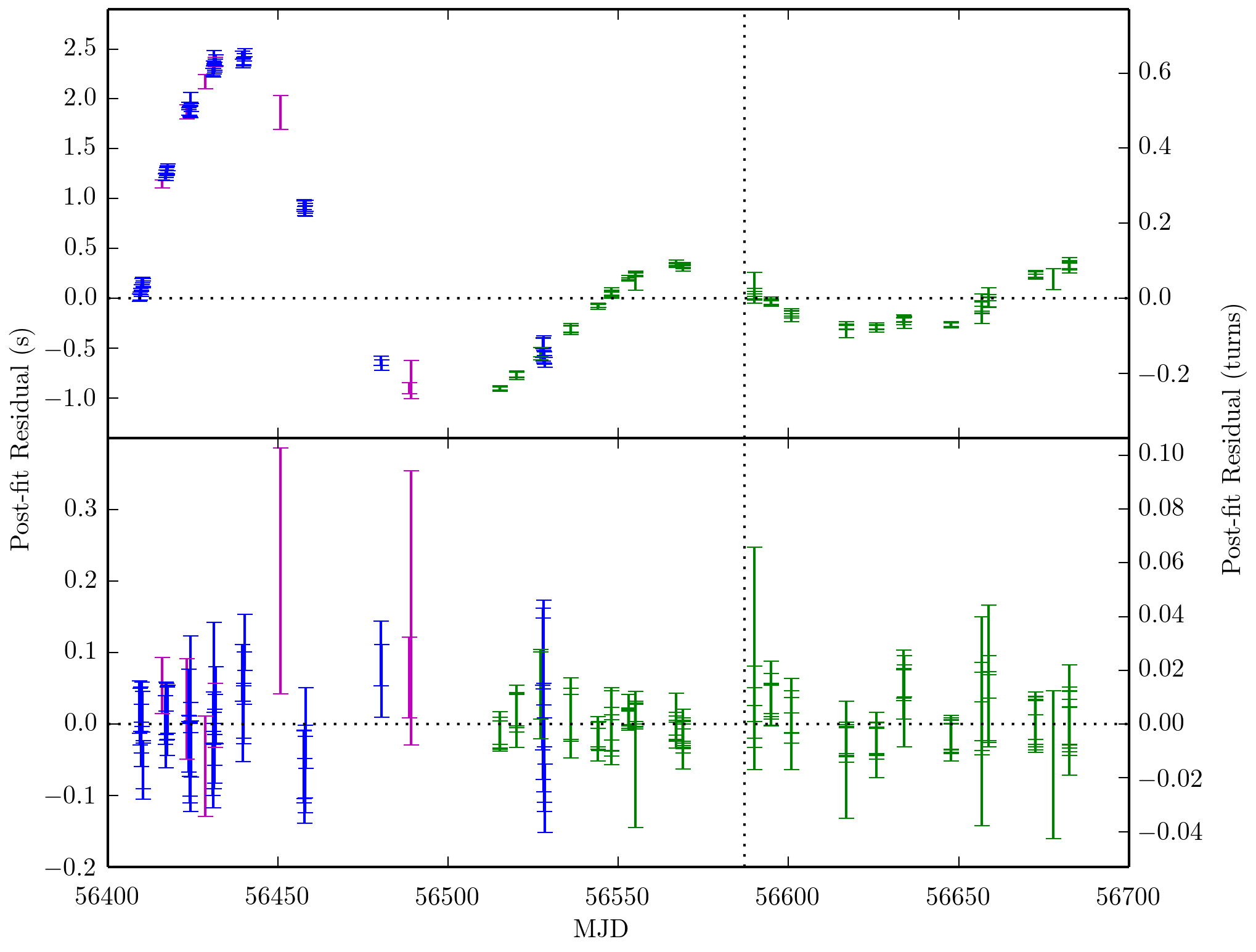}
  \caption{Post-fit timing residuals as a function of MJD.  Only data
    from the stable state are shown here because we could not
    unambiguously phase connect data from the erratic state.  TOAs
    from our GBT observations (green) are shown along with those
    previously published by \citet{kab+14} from \textit{Swift}
    (magenta) and \textit{NuSTAR}.  The vertical dotted line indicates
    the reference epoch used in our timing model.  The radio and X-ray
    TOAs are independently phase connected, with an arbitrary phase
    offset applied to align the two data sets.  \emph{Top}: The
    residual structure fitting only for $f$, $\dot{f}$, $\ddot{f}$,
    and $f^{(3)}$.  The effects of timing noise are clearly evident.
    \emph{Bottom}: The results of fitting for twelve frequency
    derivatives presented.  Note the much smaller horizontal
    scale.  \label{fig:residuals}}
\end{figure*}

Integrated radio pulse profiles of \srcnm\ are shown in Figure
\ref{fig:profiles} for each of our observing epochs.  The profile
underwent significant changes over the course of our observations.  In
our earliest data, the profile appeared double peaked, although the
two peaks were not always clearly separated prior to MJD 56544.
During this period the profile was well fit using a two-component
Gaussian model.  The most obvious changes in the profile were
variations in the relative amplitude of the two Gaussian components
from epoch to epoch, which may be due to pulse-to-pulse variability.
However, there were also more subtle, long-term changes in the
profile.  We characterized these by measuring the full width at $20\%$
peak amplitude ($W_{20}$) and the separation between the two peaks in
the profile ($\Delta$).  Each quantity was measured by fitting an
analytic, two-component Gaussian model to the on-pulse region.  The
peak amplitude was measured by finding the global maximum of the fit,
and $W_{20}$ was found using a bisection method to determine the phase
at which the model crossed the $20\%$ flux level on both the leading
and trailing sides of the pulse.  To measure $\Delta$ we simply
calculated the pulse phase corresponding to each profile peak.

Uncertainties were quantified following the method outlined by
\citet{fsk+13}.  We randomly removed half the profile bins from the
on-pulse region, fit a new Gaussian model, and re-calculated the shape
parameters as above.  We performed 1000 independent Monte Carlo trials
and calculated the mean values and the $68.27\%$ (i.e., $1$-$\sigma$)
confidence intervals for each parameter from the resulting
distributions.  The results of this analysis are shown in Figure
\ref{fig:shapes}.

There was an obvious increase in both $W_{20}$ and $\Delta$ between
MJDs 56544 and 56594.  Although there is significant scatter compared
to our uncertainties, we fit a linear trend to the data over this span
and find a rate of change in both $W_{20}$ of and $\Delta$ of $0.08
\pm 0.04^\circ\; \mathrm{day^{-1}}$.  A linear fit to the data between
MJDs 56594 and 56682 is consistent with no steady change in either
quantity.  Because both $W_{20}$ and $\Delta$ changed by the same
amount, the increase in pulse width can be attributed solely to an
increase in the peak separation.

With the onset of the erratic state, the profile changed dramatically.
Due to these large variations we did not calculate $W_{20}$ and
$\Delta$ for this period.  A third component emerged separated by
$\sim 0.1$ turns from the centroid of the persistent double-peaked
main components.  Between MJDs 56726 and 56856 the relative amplitude
of this third component varied significantly, from nearly equal to the
two other components to only barely above the noise level.
Furthermore, the primary profile component varied between the
previously described double-peaked form and a single, broad shape.

\subsection{Pulsar Timing}
\label{sec:timing}

We began the timing analysis of the radio data by fitting a noise-free
Gaussian template to a high signal-to-noise (\sn) integrated profile.
We used a the same template for data obtained during the stable state,
i.e. when the magnetar had only two profile components.  As noted
above, the profile varied from epoch to epoch during this time, which
could introduce a bias to our TOAs.  However, the profile variations
were subtle, with the total change in $\Delta$ amounting to only $\sim
0.02$ turns, which is of the same order as the RMS scatter in our
timing residuals and much less than the accumulated phase drift due to
timing noise.  As such, we are confident that the use of a single
template is sufficient for this span of observations.  The onset of
the erratic state made timing difficult, as discussed below.  In all
cases, pulse times of arrival (TOAs) were calculated via Fourier
domain cross-correlation \citep{tay92} of the templates with the
calibrated and RM corrected pulse profiles at each observing epoch.
We usually obtained one topocentric TOA per 100 rotations of the
magnetar, summing all frequency channels.

We were also able to include previously published TOAs obtained with
\textit{Swift} and \textit{NuSTAR} (see \citealt{kab+14} for a
description of the data set and how these TOAs were obtained).  The
XRT PC-mode data used to measure $2$--$10\; \keV$ flux lacked the time
resolution to measure pulsed emission, hence no TOAs are available
from this dataset.  We allow for an arbitrary phase shift between the
X-ray TOAs and those obtained with the GBT, which can absorb
differences in the profiles, phase offsets between the radio and X-ray
emission, DM delays, instrumental shifts, etc.  Hence, we cannot
determine an absolute phase offset between the radio and X-ray pulses.

\texttt{TEMPO} was used to fit a phase-coherent spin-down model to the
radio and X-ray TOAs, making use of the DE421 Solar System ephemeris
for barycentering and the TT(BIPM12) clock correction chain.  We held
the coordinates of \srcnm\ fixed at those reported by \citet{rep+13}
and the DM fixed at the valued reported by \citet{efk+13}, and hence
only fitted for the rotational frequency, $f$, and its derivatives.

We were able to unambiguously phase connect all radio TOAs during the
stable state.  We found that a simple spin-down model accounting only
for $f$ and $\dot{f}$ was insufficient for describing the long-term
rotational behavior of \srcnm\ (see the top panel of Figure
\ref{fig:residuals}) up to MJD 56682.  This is common among magnetars,
which exhibit a large degree of timing noise and typically require
many higher-order frequency derivatives to accurately model their
rotational phase.  We used twelve frequency derivatives, the maximum
allowed by \texttt{TEMPO}, but we caution that this model is not
predictive, and is only used to whiten the residuals.  Our final
solution is presented in Table \ref{table:params}.  Fully whitened
post-fit residuals are shown in the bottom panel of Figure
\ref{fig:residuals}.  The reduced $\chi^2$ of our timing solution was
large even after fitting all twelve frequency derivatives.  To obtain
a reduced $\chi^2$ of one, we multiplied the individual TOA errors by
a constant error factor, $\epsilon = \sqrt{\chi^2/\mathrm{d.o.f.}}  =
6.4$.

\begin{deluxetable}{lr}[t]
  \centering
  \tabletypesize{\footnotesize}
  \tablewidth{0pt}
  \tablecolumns{2}
  \tablecaption{Timing Parameters of \srcnm \label{table:params}}
  \startdata
  \cutinhead{Data, Statistics, \& Assumptions}
  Data Span (MJD) \dotfill & 56409--56682 \\
  $N_\mathrm{TOA}$ \dotfill & 165 \\
  Residual RMS ($\ms$) \dotfill & 25.6 \\
  Solar System Ephemeris \dotfill & DE421 \\
  Clock Correction Procedure \dotfill & TT(BIPM12) \\
  \cutinhead{Fixed Quantities}
  Right Ascension (J2000) \dotfill & $17^\mathrm{h}45^\mathrm{m}40\fs169$ \\
  Declination (J2000) \dotfill & $-29^\circ00\arcmin29\farcs84$ \\
  $\dm$ ($\dmu$) \dotfill & 1778 \\
  Reference Epoch (MJD) \dotfill & 56587.0 \\
  \cutinhead{Measured Quantities}
  $f$ ($\Hz$) \dotfill & $0.2656936554(12)$ \\
  $\dot{f}$ ($\Hz\, \s^{-1}$) \dotfill & $-1.2399(15)\times10^{-12}$ \\
  $\ddot{f}$ ($\Hz\, \s^{-2}$) \dotfill & $-1.047(13)\times10^{-19}$ \\
  $f^{(3)}$ ($\Hz\, \s^{-3}$) \dotfill & $6.7(1.9)\times10^{-27}$ \\
  $f^{(4)}$ ($\Hz\, \s^{-4}$) \dotfill & $3.74(18)\times10^{-32}$ \\
  $f^{(5)}$ ($\Hz\, \s^{-5}$) \dotfill & $-7.1(2.0)\times10^{-39}$ \\
  $f^{(6)}$ ($\Hz\, \s^{-6}$) \dotfill & $-3.90(26)\times10^{-44}$ \\
  $f^{(7)}$ ($\Hz\, \s^{-7}$) \dotfill & $-5.0(1.4)\times10^{-51}$ \\
  $f^{(8)}$ ($\Hz\, \s^{-8}$) \dotfill & $3.34(30)\times10^{-56}$ \\
  $f^{(9)}$ ($\Hz\, \s^{-9}$) \dotfill & $1.83(11)\times10^{-62}$ \\
  $f^{(10)}$ ($\Hz\, \s^{-10}$) \dotfill & $-1.37(17)\times10^{-68}$ \\
  $f^{(11)}$ ($\Hz\, \s^{-11}$) \dotfill & $-1.71(15)\times10^{-74}$ \\
  $f^{(12)}$ ($\Hz\, \s^{-12}$) \dotfill & $-5.05(43)\times10^{-81}$ \\
  \cutinhead{Derived Quantities}
  $B_\mathrm{s}$ ($\gauss$) \dotfill & $2.6018(16)\times10^{14}$ \\
  $\dot{E}$ ($\erg\, \s^{-1}$) \dotfill & $1.3005(16)\times10^{34}$ \\
  $\tau_\mathrm{c}$ ($\yr$) \dotfill & $3395.2(4.1)$ \\
  \enddata
  \tablecomments{Numbers in parentheses represent $1$-$\sigma$
    uncertainties in the last digits as determined by \texttt{TEMPO},
    scaled such that the reduced $\chi^2$ equals one.}
\end{deluxetable}

Although we could not phase connect the radio and X-ray TOAs, we were
able to find single a solution that adequately fit this entire data
set, allowing only for the phase offset between the two frequencies.
\citet{kab+14} reported a possible abrupt change in $\dot{f}$ around
MJD 56450, which suggestively was coincident to an X-ray burst.  Our
results indicate that the suggested change in $\dot{f}$ is consistent
with timing noise, and that a second distinct rotational ephemeris is
not needed after MJD 56450.  It is common to characterize timing noise
as the cumulative contribution over the span of observations of the
cubic term in the Taylor expansion of rotational phase
\citep[e.g.][]{antt94}:
\begin{equation}
\Delta_\mathrm{t}(t) = \left ( \frac{1}{6 f} |\ddot{f}| t^3 \right)
\end{equation}
where $t$ is the duration of timing observations.  We find
$\Delta_\mathrm{t} = 864\; \s$ (230 cycles) over a time span of
approximately 273 days.  This is significantly larger than in other
radio magnetars: $\sim 120\; \s$ (22 cycles) over 277 days for
XTE~J1810$-$197 \citep{ccr+07}, $\sim 124\; \s$ (60 cycles) over $6$
months for 1E~1547.0$-$5408 \citep{crj+08}, and $\sim 1080\; \s$ (250
cycles) over $20$ months for PSR~J1622$-$4950 \citep{lbb+12}.
However, if we do not include the X-ray TOAs and instead restrict our
analysis to the period covered by our phase-connected GBT
observations, we find $\Delta_\mathrm{t} = 140\; \s$ (37 cycles) over
a time span of approximately 167 days.  This is closer to the level of
timing noise observed in other radio magnetars.

The onset of the erratic state and the accompanying profile changes
required a change in our timing analysis, as the simple standard
template we used during the stable state was no longer adequate.  We
explored three options for obtaining TOAs: using the same
three-component Gaussian template for all epochs, using a different
Gaussian template for each epoch, and using a template based on the
folded profiles at each epoch, but with noise removed using a wavelet
smoothing algorithm.  In the latter two cases, we attempted to align
the different templates using the peak of the leading profile
component as our reference point.  We obtained TOAs using each method
and attempted to extend the solution obtained during the stable state,
but could not unambiguously maintain phase connection.  We also tried
to use a subset of TOAs from epochs just before the onset of the
erratic state to establish a new solution without the need for many
higher order frequency derivatives, and then to extend this simplified
solution into the erratic state.  Again, we could not unambiguously
phase connect the data.  Finally, we attempted to form a solution from
only the TOAs obtained during the erratic state, as the magnetar could
have experienced a glitch or other sudden change in rotational
parameters, but we still could not obtain a phase-connected solution.
We are therefore unable to provide a precise timing solution for the
erratic state.  The implications of this are discussed in
\S\ref{sec:discussion}.

\subsection{Single Pulses}
\label{sec:singlepulse}

\subsubsection{Energy distribution}
\label{sec:spenergy}

\begin{figure}[t]
  \centering
  \includegraphics[width=\columnwidth]{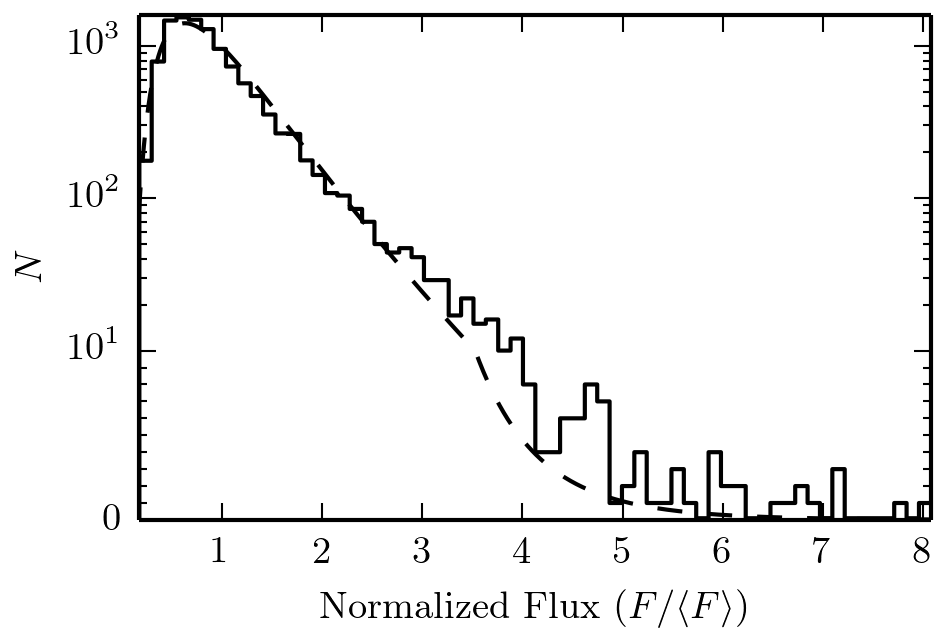}
  \caption{A histogram of peak single-pulse flux, normalized by the
    mean peak flux at that epoch.  The dashed line shows the best-fit
    log-normal distribution, as described in the text.  The presence
    of a high-energy tail is clearly visible above four times the mean
    flux.  Note that the ordinate scaling changes from linear to
    logarithmic at values above 10.  \label{fig:sphist}}
\end{figure}

\srcnm\ emits bright single pulses during most rotations.  We analyzed
the energy distribution, sub-pulse structure, and emission phase of
the pulses.  We recorded the peak flux in the on-pulse region during
each rotation, but discarded a rotation if the peak flux never reached
$\geq 3$ times the off-pulse noise level.  The peak fluxes at each
epoch were normalized by the mean peak flux at that epoch.  A
histogram of these normalized fluxes is shown in Figure
\ref{fig:sphist}.  The distribution of peak fluxes roughly follows a
log-normal distribution with a logarithmic mean $\mu \approx -0.163$
and standard deviation $\sigma \approx 0.548$.  There appears to be a
high-energy tail to the observed distribution, however.  The
log-normal approximation underestimates the observed number of pulses
at fluxes $F \gtrsim 2.5 \langle F \rangle$, and especially at $F
\gtrsim 4 \langle F \rangle$.  We note, though, that only 416 pulses
out of 11428 (about $3.6\%$) have $F \geq 2.5 \langle F \rangle$.
\citet{lbb+12} found that the single pulses of PSR~J1622$-$4950 also
followed a log-normal distribution.  A careful analysis of single
pulses from XTE~J1810$-$197 by \citet{ssw+09} found a more complex
distribution of single pulse energies, with emission sometimes best
described by a combination of a power law and log-normal distribution
because of the presence of a high-energy tail.  In this regard,
\srcnm\ seems similar to XTE~J1810$-$197.

\subsubsection{Drifting Sub-pulses}
\label{sec:subpulses}

We employed the 2D fluctuation spectrum method of \citet{es02} to
search for and characterize any potential drifting sub-pulses.  This
method relies on a power spectrum of the 2D Fourier transform of pulse
flux as a function of pulse phase and pulse number, i.e.
\begin{equation}
S(u,v) = \left | \frac{1}{K} \sum_{j=0}^{n_\mathrm{bins}-1} 
                 \sum_{k=0}^{n_\mathrm{pulses}-1} F(j,k)\: e^{-2 \pi i (uj + vk)}
         \right |^2,
\end{equation}
where $K = n_\mathrm{bins} \times n_\mathrm{pulses}$ is a
normalization factor \citep{lk05}.  The signature of drifting
sub-pulses is harmonic structure in $S(u,v)$, from which we can
determine the characteristic spacing between sub-pulses and the period
with which sub-pulses drift in phase.  

The baseline fluctuations described in \S\ref{sec:data} caused
significant red noise in $u$.  There was sometimes an increase in
power in $v$, but this was seen even when we only analyzed the
off-pulse region (either in full or a randomly selected subset).  As
such, we attribute it to an artifact of the data processing.  We see
no evidence for drifting sub-pulses in \srcnm.

\subsection{Search for Off-pulse Radio Burst Emission}
\label{sec:spoffpulse}

Some magnetars are observed to emit extremely bright X-ray bursts
lasting $\sim \mathrm{few}~\ms$.  These bursts are sporadic and can
occur at phases not typically associated with the on-pulse region.
The magnetic reconnection model proposed by \citet{lyu02} makes a
clear prediction that such X-ray bursts should be accompanied by
simultaneous radio bursts with very high fluxes.  This model is
particularly interesting in light of the recent discovery of a
population of extremely bright, short duration radio bursts of
apparently cosmological origin \citep{lbm+07,tsb+13}.  Unfortunately,
we were unable to obtain simultaneous X-ray/radio observations of
\srcnm, but we can still search for radio bursts that occur outside of
the on-pulse region which, if detected, would be highly suggestive.
For this analysis, we used tools in the
\texttt{PRESTO}\footnote{\url{http://www.cv.nrao.edu/~sransom/presto/}}
software suite \citep{rem02} that do not require identifying an
on-pulse region.  We de-dispersed the raw GBT data using an RFI mask
generated by \texttt{PRESTO} at both the DM of the magnetar ($1778\;
\dmu$) and at $\dm = 0\; \dmu$, so that we could reject any remaining
RFI.  We then conducted a blind search for single pulses by
match-filtering the data with boxcar functions of various widths,
ranging from $163.84\; \us$ (our native resolution) to $\sim 5\; \ms$.
We then calculated the corresponding pulse phase of all single pulses
with $\sn \geq 5$, rejecting pulses that also appeared in the
un-dispersed time series.  No pulses were detected outside of the main
pulse window.  Of course, this does not rule out the \citet{lyu02}
model, since it may simply be the case that no X-ray bursts occurred
during our observations.

\section{Discussion}
\label{sec:discussion}

\srcnm\ is similar to other radio magnetars in many regards.  It has a
high degree of linear polarization, with a lower but significant
fraction of circular polarization.  It exhibits a high degree of
timing noise.  Its single pulse energy distribution is similar to that
of XTE~J1810$-$197.  However, during the stable state it showed a
level of radio flux density and profile stability not often seen in
radio magnetars.

During the stable state, we observed a steady evolution in the
separation between the two profile peaks.  Prior to the start of our
GBT observations, the profile was apparently single-peaked.  The
growing separation between the peaks that we observed may suggest that
the profile evolved smoothly from single to double peaked.  However,
the extended periods of minimal change in peak separation that we
observed indicate that any such smooth change did not occur steadily.
The emergence of the third profile component marked the onset of the
erratic state, but as noted in \S\ref{sec:shape}, there are hints of a
widely separated component in some of the profiles presented by
\citet{efk+13}.  A joint analysis of both data sets could link this
with the third component we observed.  If this were the case it would
raise the question of why the component remained dormant for many
months before abruptly reappearing.  

Unlike the radio flux density, the X-ray flux has decayed mostly
steadily since the initial outburst, although some short bursts have
been detected with the \textit{Swift} BAT (see \S\ref{sec:overview}).
For comparison, the radio magnetar XTE~J1810$-$197 faded in both the
radio and X-ray bands following its discovery \citep{ccr+07}.  The
X-ray flux of 1E~1547$-$5408 decayed steadily following an outburst,
while the radio flux density varied \citep{frj+08}.  PSR~J1622$-$4950
was discovered via its radio emission while in an X-ray quiescent
state, with no X-ray outbursts detected in archival data as far back
as four years prior to its radio discovery \citep{lbb+10}.  This
suggests that the radio emission in PSR~J1622$-$4950 is long-lived and
not related to its X-ray emission, and in this regard appears similar
to \srcnm.

What caused \srcnm\ to move into the erratic state?  The change in
mean flux and profile morphology suggest a significant reconfiguration
of the magnetosphere.  It would be useful to know if this was
associated with a glitch or sudden change in spin-down, as is
sometimes correlated with profile changes in other radio pulsars
\citep{lhk+10}.  Unfortunately, our inability to maintain phase
connection during this state prevents us from making such a definitive
statement.  We can speculate on the reasons for losing phase
connection.  One possibility is that the profile changes introduced
too much uncertainty into the definition of a fiducial point in the
profile to maintain phase connection.  It is also possible that the
gap between our observations was simply too large to maintain
phase-coherence, given the uncertainties in the rotational parameters.
We cannot rule this out, but we do note that the timing model
presented by \citet{kab+14} was accurate enough to maintain phase
connection throughout the entire stable state, even before making any
adjustments to the solution based on the radio data.  Timing noise was
clearly evident, but phase connection was maintained.  The duration of
the stable state was far greater than the $\sim 30$ day gap between
the observations over which we lost phase connection, and introducing
an artificial $\sim 30$ day gap in our data does not result in a lost
of phase connection during the stable state.  On the other hand, based
on our analysis of timing noise presented in \S\ref{sec:timing},
\srcnm\ exhibited a degree of timing noise unprecedented in radio
magnetars when we included the time span covered by \textit{NuSTAR}
and \textit{Swift} timing observations.  This timing noise was
apparently large enough that it could also be interpreted as a sudden
change in $\dot{f}$, possibly associated with an X-ray burst
\citep{kab+14}.  Thus, there is precedent for timing noise in
\srcnm\ sufficient to introduce ambiguities in phase connected timing
solutions.  It is interesting to note, however, the coincidence
between the radio variability and our inability to phase-connect, with
both beginning roughly $\sim 300$ days after the initial outburst.  If
our timing difficulties are not due to the gap or to the varying pulse
profile, then this is reminiscent of behavior seen in magnetar 1E
1048.1$-$5937, in which, now three times, delayed torque variability
followed, after a $\sim 100$ day decay, an X-ray outburst
\citetext{Arcibald et al. submitted}.  In the latter case, radio
emission is not seen, possibly due to unfavorable beaming.  We
speculate that we could therefore be seeing similar behavior in
\srcnm, but with the additional radio diagnostic.  Under this scenario
we would expect a relatively stable X-ray pulse profile for
\srcnm\ during the radio-variable phase.  Unfortunately, we cannot
verify this due to the low XRT PC-mode time resolution.

It is tempting to speculate that the increase in $44$-$\GHz$ flux
density \citep{yrh+14} is connected with the emergence of the third
profile component.  Unfortunately, the gap between the high frequency
brightening and our closest observations prevent us from associating
these events definitively.  The measured flux of both observations
that were taken closest to the $44$-$\GHz$ brightening is not
anomalously high, but we do see an eventual increase in
$S_\mathrm{8.7\; GHz}$.  This suggests that the $44$-$\GHz$
brightening was not long-lived, but may be associated with greater
variability in flux overall.

\section{Conclusions}
\label{sec:conclusions}

We observed the magnetar \srcnm\ for eleven months using the GBT,
measuring its radio flux density, pulse profile shape, single-pulse
behavior, and timing parameters.  We have also analyzed publicly
available \textit{Swift} XRT data from the initial outburst of the
magnetar, tracking its X-ray flux and spectral evolution.  We find
that for the first 5.5 months of our GBT observations, the radio flux
density and pulse profile remained relatively stable, with a slow
increase in the pulse width and separation between two profile peaks.
This is in contrast to the three other radio magnetars, which were all
highly variable in the radio band.  During this time, the magnetar
exhibited a high degree of timing noise but did not otherwise
experience any anomalous rotational behavior.  After this stable
period, \srcnm\ entered an erratic state marked by a higher and more
variable radio flux density and significant changes in the radio pulse
profile from epoch to epoch.  We were unable to maintain phase
connection, but can only speculate as to the causes.  The onset of
this erratic state occurred within two weeks of a short-lived increase
in radio density at $44\; \GHz$ measured with the VLA \citep{yrh+14}.
The X-ray flux of \srcnm\ has steadily decayed since the initial
outburst, and did not deviate from this trend at any point during the
time span covered by our GBT observations, including during the
erratic radio state.  We conclude that whatever caused the erratic
radio state is decoupled from the X-ray emission.

\acknowledgments

The authors wish to thank Paul Demorest for invaluable assistance in
processing the GBT data, and for helping to track down and correct
some instrumental and hardware errors.  The GBT is operated by the
National Radio Astronomy Observatory, which is a facility of the
National Science Foundation operated under cooperative agreement by
Associated Universities, Inc.  We acknowledge the use of public data
from the \textit{Swift} data archive.  This research made use of the
XRT Data Analysis Software (XRTDAS) developed under the responsibility
of the ASI Science Data Center (ASDC, Italy).  The \textit{NuSTAR}
mission is a project led by the California Institute of Technology,
managed by the Jet Propulsion Laboratory, and funded by the National
Aeronautics and Space Administration.  This research made use of the
\textit{NuSTAR} Data Analysis Software (NuSTARDAS) jointly developed
by ASDC and the California Institute of Technology.  V.\ Kaspi
receives support from an NSERC Discovery Grant and Accelerator
Supplement, from the Centre de Recherche en Astrophysique du
Qu\'{e}bec, an R. Howard Webster Foundation Fellowship from the
Canadian Institute for Advanced Study, the Canada Research Chairs
Program, and the Lorne Trottier Chair in Astrophysics and
Cosmology. R.\ Archibald receives support from a Walter C. Sumner
Memorial Fellowship.

\bibliographystyle{apj}
\bibliography{references}{}

\end{document}